\documentclass[10pt]{article}
\usepackage{amssymb,epsfig}
\advance \textheight by \topskip
\def\d{\partial}
\def\l{\left(}
\def\r{\right)}
\def\la{\langle}
\def\ra{\rangle}
\def\S{{\cal S}}
\newcommand{\be}{\begin{equation}}
\newcommand{\ee}{\end{equation}}

\newcommand{\Tr}{{\rm Tr}}
\newcommand{\bg}{\begin{gather}}
\newcommand{\eg}{\end{gather}}

\def\half{\frac{1}{2}}

\begin{document}      

\title{CompHEP Package with Light Gravitino and Sgoldstinos}

\author{Dmitry~S.~Gorbunov\footnote{{\bf e-mail}: gorby@ms2.inr.ac.ru}~$^{,1}$ ~and Andrei~V.~Semenov\footnote{{\bf e-mail}: semenov@theory.sinp.msu.ru}~$^{,2,3}$\\\\
$^1${\small{\em Institute for Nuclear Research of the Russian Academy of
Sciences,}} \\
{\small{\em 60th October Anniversary prospect 7a, Moscow 117312, Russia}}\\
$^2${\small{\em Laboratoire de Physique Theorique LAPTH,}} \\
{\small{\em Chemin de Bellevue, B.P. 110, F-74941, Annecy-le-Vieux, 
Cedex, France}}\\
$^3${\small{\em Laboratory of Particle Physics, Joint Institute for Nuclear Research,}} \\
{\small{\em Dubna 141980, Moscow region, Russia}}}
\date{}
\maketitle

\begin{abstract}
We present a complete description of a new physical model with goldstino supermultiplet for CompHEP software package. The model contains MSSM content supplemented by three new fields: {\it goldstino} -- two component spinor, scalar {\it sgoldstino} and pseudoscalar {\it sgoldstino}. To the leading order in $1/F$, where $F$ is the parameter of supersymmetry breaking, and to zero order in MSSM gauge and Yukawa coupling constants, the interactions between the component fields of goldstino supermultiplet and MSSM fields are derived. They correspond to the most attractive for collider studies processes where only one of these {\it new} particles appears in a final state. In this case light gravitino behaves exactly as goldstino. For sgoldstinos, as they are R-even, we include in the package only sgoldstino couplings to goldstino and sgoldstino couplings to SM fields as the most interesting phenomenologically. All new coupling constants between the components of goldstino supermultiplet and MSSM fields are completely determined by the ratios of soft supersymmetry breaking parameters and $F$. Thus, in addition to the MSSM coupling constants there are three new parameters in the package: the scale of supersymmetry breaking, $\sqrt{F}$, and masses of scalar and pseudoscalar sgoldstinos. The CompHEP notations for the new fields and parameters are presented. The package was successfully tested and all relevant results known in literature have been reproduced with good accuracy. It may be used in order to calculate any tree-level process with one gravitino or sgoldstino on mass shell. Universality of the Lagrangian allows to apply the package in study of phenomenology of any supersymmetric extension of the Standard Model. 
\end{abstract}

\section{Introduction}
CompHEP package~\cite{Pukhov:1999gg} is adjusted to calculate automatically particle decay rates and cross sections at the tree level.\footnote{The total number of particles in initial and final states does not exceed 6.} It is aimed to improve the accuracy, to cut down the efforts and to shorten the time usually required in study of the collision processes at high energy. 

First versions of CompHEP contained physical models based only on 
the set of particles and interactions forming the Standard Model (SM). 
In CompHEP any user has the opportunity to change vertices and add 
new particles into the SM content. Thereby some 
modified models (based on the SM) have been created and used in
calculations (see e.g. Refs.~\cite{Ilin:1995jv} for leptoquark 
models and Refs.~\cite{Boos:2001sj} for SM with anomalous couplings). 
The growth of activity in a study of phenomenology of various extensions of the SM inspired the authors of the package to include the Minimal Supersymmetric extension of the Standard Model (MSSM) into CompHEP.\footnote{The package is freely available at {\bf URL:http://theory.sinp.msu.ru/\~{}semenov/mssm.html}} Below we will refer to this model as {\it CompHEP/SUSY} one. 

In the framework of supersymmetry there are several types of models which give significantly different predictions for the values of low-energy MSSM parameters and for low-energy spectrum, that might be tested at accelerators. The incorporation of different classes of models into CompHEP gives an opportunity to study quantitatively high energy phenomenology of these models getting results with high accuracy and with minimum boring routines. 
The first step to this direction had been done, when the possibility to work with the set of parameters evaluated in the frameworks of mSUGRA and GMSB became available in CompHEP/SUSY. The incorporation of other extensions, in particular, the MSSM with R-parity violation, are in progress. 

This paper is devoted to the description of the supersymmetric extensions of the SM with very light {\it gravitino} and light {\it sgoldstinos}. These new particles as well as their interactions with MSSM fields have been included in the new version of CompHEP/SUSY model.\footnote{The package is freely available at {\bf URL:http://theory.sinp.msu.ru/\~{}semenov/mssm.html}} The paper is organized as follows. In section 2 we describe the supersymmetric model with sgoldstino supermultiplet. Namely, section 2.1 contains the description of goldstino supermultiplet. In sections 2.2 and 2.3 we present the effective Lagrangians for very light gravitino and light sgoldstinos, respectively. In section 2.4 we derive the free Lagrangian for light gravitino and sgoldstino fields, as well as the effective interactions responsible for sgoldstino decays into gravitinos. A sketch of phenomenology of the models with very light gravitino and models with light sgoldstinos is presented in Section 3. It is aimed to outline the experimentally allowed region of parameter space of the model. Section 4 explains the CompHEP notations for the new particles and parameters additional to ones used in CompHEP/SUSY model. Section 5 is devoted to concluding remarks, the tests which have been done to check the incorporated models, and possible ways to improve the package.   
  
\section{Description of the model}

\subsection{Goldstino supermultiplet}
In any supersymmetric extension of the Standard Model (SM) of particle
physics spontaneous supersymmetry breaking occurs due to 
non-zero vacuum expectation value of an auxiliary component 
of some chiral or vector superfield. 
As the most simple case, let us consider a model where the only 
chiral superfield
\begin{equation}
\S=s+\sqrt{2}\theta\psi+\theta^2F_s
\label{spurion}
\end{equation}
obtains non-zero vacuum expectation value $F$ for the auxiliary
component, 
\begin{equation}
\la F_s\ra\equiv F\;.
\label{spurion-1}
\end{equation}
Then $\psi$ is two-component Goldstone fermion, {\it goldstino},  
and its superpartners 
\begin{equation}
S\equiv\frac{1}{\sqrt{2}}(s+s^*)\;,~~~~
P\equiv\frac{1}{i\sqrt{2}}(s-s^*)\;,
\label{definition}
\end{equation}
are scalar and pseudoscalar {\it sgoldstinos}, respectively. 

In the framework of supergravity $\partial_\mu\psi$ becomes a longitudinal
component of gravitino, that is the super-Higgs effect. 
As a result, gravitino gains mass $m_{3/2}$ 
which is completely determined, in a realistic model with zero 
cosmological constant, by the parameter of
supersymmetry breaking $F$: 
\begin{equation}
m_{3/2}=\frac{\sqrt{8\pi}}{\sqrt{3}}\frac{F}{M_{Plank}}\;.
\label{gravitino-mass}
\end{equation} 
Sgoldstinos remain massless at tree level and become massive due to corrections from high order terms in the K\"ahler potential. Provided these terms are
sufficiently suppressed sgoldstinos are light and may appear in particle collisions at high energy colliders. Such pattern emerges in a number of non-minimal supergravity models~\cite{ellis} and also in gauge mediation
models if supersymmetry is broken via non-trivial superpotential (see, 
Ref.~\cite{gmm} and references therein). Here we consider sgoldstino masses as free parameters. 

Gravitino and sgoldstinos interact with MSSM fields and corresponding coupling constants are inversely proportional to the parameter of supersymmetry breaking $F$. This implies that gravitino has to be very light (see Eq.~(\ref{gravitino-mass})), otherwise gravitino as well as sgoldstinos is effectively decoupled from the MSSM fields at the energy scale which is expected to be available for study at the nearest future colliders.

\subsection{Gravitino effective Lagrangian}

The effective Lagrangian for the gravitino $\tilde{G}_\mu$ is obtained from the N=1 supergravity theory~\cite{Cremmer:1982en}, and may be used to calculate scattering processes involving all helicity components of the massive gravitino. Meanwhile the energy scale attainable at the present and the nearest future generation of accelerators favors the longitudinal component of gravitino as the most promising to be searched for in collision experiments. The matter is that while the effective coupling constants for $\pm{3\over2}$ helicity components of gravitino are suppressed by gravity scale, the longitudinal component of gravitino, in accordance with the Equivalence Theorem~\cite{Casalbuoni:1988kv}, effectively behaves as a two-component fermion, {\it goldstino} $\psi$, 
\begin{equation}  
\tilde{G}_\mu\sim i{\sqrt{2}\over\sqrt{3} m_{3/2}}\d_\mu\psi\;,
\label{equivalence}
\end{equation}
so gets enhancement if the energy $E$ of the process 
is significantly larger than gravitino mass. 
Therefore, goldstino couplings are suppressed by the ratio $E/F$ 
(see Eq.~(\ref{gravitino-mass})), which may be 
significantly lower than gravity scale. The effective low energy Lagrangian 
is non-renormalizable, and on general grounds the sensitivity of collider 
experiments to the scale of supersymmetry breaking $\sqrt{F}$ is expected to 
be not better than about 10 TeV. Thus, in the models under discussion 
gravitino is very light, $m_{3/2}\lesssim{a~few}\cdot10^{-2}$~eV (lighter 
than a few tens millielectronvolts), and only longitudinal component, goldstino, is relevant. Hereby gravitino is considered as massless ($M_{Plank}\to\infty$, 
$F=fixed$) in this approach. 

Further simplification is related to the fact that cross sections of the processes with few goldstinos in a final state are suppressed by additional powers of $F$ in comparison with single goldstino production. Consequently, searches for the latter process are expected to be more sensitive to the scale of supersymmetry breaking, so we retain only one-goldstino couplings in the effective Lagrangian. In this case the interaction between goldstino and other fields is plainly given by Goldberger--Treiman formula
\[
{\cal L}_{GT}=\frac{1}{F}J^{\mu}_{ SUSY}\partial_\mu\psi\;,
\] 
with $J^{\mu}_{SUSY}$ being a supercurrent. For MSSM this interaction reads
\begin{eqnarray}
{\cal L}_\psi&=&-{1\over\sqrt{2}F}\cdot\!\!\!\!\sum_{all~gauge\atop fields}\!\!\!\!\l
\d_\mu\psi\sigma^\nu\bar{\sigma}^\rho\sigma^\mu\bar{\lambda}^\alpha_a-
\d_\mu\bar{\psi}\bar{\sigma}^\nu\sigma^\rho\bar{\sigma}^\mu\lambda^\alpha_a
\r F_{a~\nu\rho}^\alpha
\nonumber
\\&+&
{i\over F}\cdot\!\!\!\!\!\!\!\!\sum_{all~matter\atop and~Higgs~fields}
\!\!\!\!\l\l  
D_\nu\tilde{\phi}_k\r^\dag\d_\mu\bar{\psi}\sigma^\nu\bar{\sigma}^\mu\phi_k
-\bar{\phi}_k\bar{\sigma}^\mu\sigma^\nu\d_\mu\bar{\psi}D_\nu\tilde{\phi}_k 
\r\;,
\label{goldstino-mssm}
\end{eqnarray}
where $\alpha$ and $k$ run over component fields of vector 
$(\lambda_a^\alpha,~\!\upsilon_{a~\mu}^\alpha)$ and 
chiral $(\tilde{\phi}_k,~\!\phi_k)$  MSSM supermultiplets, respectively 
(see Eq.~(\ref{notation}) for details); 
$\sigma^\mu=(1,\vec{\sigma})$, 
$\bar{\sigma}^\mu=(1,-\vec{\sigma})$. 
Throughout this paper we adopt two-component fermions.   

Whereas we are interested in processes with light gravitino at an external line of Feynman graphs (i.e., tree-level production of gravitino), we simplify Eq.~(\ref{goldstino-mssm}) by making use of the goldstino equation of motion (equation of motion of a free massless two-component spinor field). Finally we obtain
\begin{equation}
{\cal L}_\psi={\sqrt{2}\over F}\cdot\!\!\!\!\sum_{all~gauge\atop fields}\!\!\!\!
F_{a~\nu\rho}^\alpha\l\d^\rho\bar{\psi}\bar{\sigma}^\nu\lambda_a^\alpha
-\d^\rho\psi\sigma^\nu\bar{\lambda}_a^\alpha\r
+
i{2\over F}\cdot\!\!\!\!\!\!\!\!\sum_{all~matter\atop and~Higgs~fields}
\!\!\!\!\l \l D_\mu\tilde{\phi}_k\r^\dag\phi_k\d^\mu\psi
-\d_\mu\bar{\psi}\bar{\phi}_k D^\mu\tilde{\phi}_k\r\;.
\label{goldstino}
\end{equation} 
Such interaction terms exist in any supersymmetric extension 
of the SM, so this Lagrangian may be regarded as the universal tool to study the sensitivity of various processes to the scale of supersymmetry breaking.  

It is worth noting, that as the total amplitudes with helicity $\pm\half$ gravitino in a final state have to vanish unless supersymmetry is broken, these amplitudes should be proportional to some supersymmetry breaking parameters. Therefore one can use the equations of motion of MSSM fields to rewrite the effective Lagrangian~(\ref{goldstino}) in the form, where all vertices are proportional to soft breaking terms. Although this form is very eligible for simple estimates, since the leading order dependence of the amplitudes on the soft parameters is explicit, CompHEP operates with the Lagrangian~(\ref{goldstino}) because of some technical reasons. 

\subsection{Sgoldstino effective Lagrangian}

We obtain the low-energy effective Lagrangian for sgoldstino by making use of spurion method~\cite{Brignole:1996fn}. It consists in exploiting the fact that, by definition, sgoldstinos are scalar components of the very supermultiplet $\S$ whose auxiliary component gains non-zero vacuum expectation value due to supersymmetry breaking. Then one can consider a simple supersymmetric model with non-renormalizable interactions between $\S$ and MSSM superfields, which provide MSSM soft terms if non-zero $\la F_s\ra$ is taken into account. Consequently, the corresponding coupling constants are fixed by the ratios of the soft terms and $F$. 

We begin with the Lagrangian of the MSSM, which 
consists of two parts: the first part is a minimal supersymmetric generalization of the Standard Model (SM), while the second one represents the supersymmetry breaking, 
\[
{\cal L}_{MSSM}={\cal L}_{SUSY}+{\cal L}_{breaking}\;,
\] 
where 
\[
{\cal L}_{SUSY}={\cal L}_{gauge}+{\cal L}_{K\ddot{a}hler}+{\cal L}_{superpotential}\;,
\] 
and 
\begin{eqnarray*}
{\cal L}_{gauge}={1\over 4}\sum_{all~gauge\atop fields}\l\int\!\!d^2\theta~\Tr W^\alpha W^\alpha+h.c.\r\;,\nonumber\\
{\cal L}_{K\ddot{a}hler}=\sum_{all~matter\atop and~Higgs~fields}\int\!\! d^2\theta~ d^2\bar{\theta}~\Phi_k^\dagger e^{g_1V_1+g_2V_2+g_3V_3}\Phi_k\;,
\\
{\cal L}_{superpotential}=\int\!\!d^2\theta~\epsilon_{ij}\!\l \mu H_D^iH_U^j+y^L_{ab}L_a^jE_b^cH_D^i+
y_{ab}^DQ_a^jD_b^cH_D^i+y_{ab}^UQ_a^iU_b^cH_U^j\r+h.c.\;,
\end{eqnarray*}
where $L$, $E$ and $Q$, $U$, $D$ are lepton and quark superfields, respectively, $H_D$ and $H_U$ are electroweak Higgs superfields; $i,j=1,2$ are SU(2) and $a,b=1,2,3$ are generation indices, color indices are omitted and two-component antisymmetric tensor is defined as $\epsilon_{12}=-1$; $y_{ab}$ are matrices of Yukawa constants and $\mu$ is a parameter of higgsino mixing. Hereafter we assume that R-parity is conserved. The expansion of superfields in series of grassmanian variables $\theta$,$\bar{\theta}$ introduces the component fields: 
\begin{eqnarray}
\Phi_k=\tilde{\phi}_k+\sqrt{2}\theta\phi_k+\theta^2F_k\;,\nonumber\\ 
V=T^aV_a\;,~~~~\Tr~T^aT^b=\delta^{ab}\;,~~~~V_a=-\theta\sigma^\mu\bar{\theta}\upsilon_{a~\mu}+i\theta^2\bar{\theta}\bar{\lambda}_a+\half\theta^2\bar{\theta}^2D_a\;,\nonumber\\ 
W^\alpha=T^aW^\alpha_a\;,~~~~W^\alpha_a=-i\lambda_a^\alpha+\theta D_a^\alpha-{i\over2}\sigma^\mu\bar{\sigma}^\nu\theta F_{a~\mu\nu}^\alpha+\theta^2\sigma^\mu D_\mu\bar{\lambda}^\alpha_a\;,\nonumber\\
F_{a~\mu\nu}^\alpha=\d_\mu\upsilon_{a~\nu}^\alpha
-\d_\nu\upsilon_{a~\mu}^\alpha+f_{abc}~\!
\upsilon_{b~\mu}^\alpha\upsilon_{c~\nu}^\alpha\;,~~~~
D_\mu\bar{\lambda}^\alpha_a=\d_\mu\bar{\lambda}^\alpha_a+f_{abc}~\!\upsilon_{b~\mu}^\alpha\bar{\lambda}^\alpha_c\;, 
\label{notation}
\end{eqnarray}
where for vector superfields we chose the Wess--Zumino gauge. 

Generally, supersymmetry breaking part has the form
\begin{eqnarray}
-{\cal L}_{breaking}=\sum_km_k^2|\tilde{\phi}_k|^2+\l\half
\sum_\alpha M_\alpha~\Tr\lambda^\alpha\lambda^\alpha+h.c.\r\nonumber
\\\label{soft-terms}
-\epsilon_{ij}\l 
Bh^i_Dh^j_U+
A^L_{ab}\tilde{l}_a^j\tilde{e}_b^ch_D^i+
A_{ab}^D\tilde{q}_a^j\tilde{d}_b^ch_D^i+
A_{ab}^U\tilde{q}_a^i\tilde{u}_b^ch_U^j+h.c.\r\;,
\end{eqnarray} 
where $k$ ($\alpha$) runs over all scalar $\tilde{\phi}_k$ (gaugino $\lambda_\alpha$) fields.

Since supersymmetry is broken {\it spontaneously}, Eq.~(\ref{soft-terms}) implies the following effective interaction between MSSM superfields and goldstino supermultiplet $\S$:   
\[
{\cal L}_{\S-MSSM}={\cal L}_{\S-K\ddot{a}hler}+{\cal L}_{\S-gauge}+{\cal L}_{\S-superpotential}\;,
\]
where 
\begin{eqnarray*}
{\cal L}_{\S-K\ddot{a}hler}&=&-\int\! d^2\theta~ d^2\bar{\theta}~\S^\dag\S\cdot\!\!\!\!\!\!\!\!\sum_{all~matter\atop and~Higgs~fields}{m_k^2\over F^2}~\!\Phi_k^\dag 
~\!e^{g_1V_1+g_2V_2+g_3V_3}~\!\Phi_k\;,
\\
{\cal L}_{\S-gauge}&=&{1\over 2}\int\!\!d^2\theta~
\S\cdot \!\!\!\!\!\!\sum_{all~gauge\atop fields}{M_\alpha\over F}~\!Tr W^\alpha W^\alpha+h.c.\;,
\\
{\cal L}_{\S-superpotential}&=&\int\!\!d^2\theta~\S\cdot\epsilon_{ij}\!
\l{B\over F}~\!H_D^iH_U^j+{A^L_{ab}\over F}~\! L_a^jE_b^cH_D^i+
{A_{ab}^D\over F}~\!Q_a^jD_b^cH_D^i+{A_{ab}^U\over F}~\!Q_a^iU_b^cH_U^j\r+h.c.
\end{eqnarray*}
These terms emerge if the fields from the hidden sector, where supersymmetry breaking occurs, are integrated out. The only remnant of the hidden sector is goldstino supermultiplet, which is supposed to be light.  

Integrating over $\theta$, $\bar{\theta}$ and taking into account Eqs.~(\ref{spurion}), (\ref{spurion-1}) we obtain soft supersymmetry breaking terms~(\ref{soft-terms}) as well as interaction between the components of goldstino supermultiplet and components of MSSM superfields: 
\begin{eqnarray}
{\cal L}_{\S-K\ddot{a}hler}&=&-\!\!\!\!\!\!\sum_{all~matter\atop and~Higgs~fields}\l{m_k^2\over F}s\cdot\tilde{\phi}_k^\dag F_k+h.c.\r\;,
\label{S-kahler}
\\
{\cal L}_{\S-gauge}&=&\!\!\!\!\sum_{all~gauge\atop fields}\biggl(
-i{M_\alpha\over F}s\cdot\lambda^\alpha_a\sigma^\mu D_\mu\bar{\lambda}^\alpha_a+{M_\alpha\over2F}s\cdot D^\alpha_a D^\alpha_a
\nonumber
\\&-&
{M_\alpha\over4F}s\cdot F_{a~\mu\nu}^\alpha F_a^{\alpha~\mu\nu}-i{M_\alpha\over 8F}s\cdot F_{a~\mu\nu}^\alpha \epsilon^{\mu\nu\lambda\rho}F_{a~\lambda\rho}^\alpha+h.c.\biggr)\;,
\label{S-gauge}
\\
{\cal L}_{\S-superpotential}&=&-\epsilon_{ij}\biggl( 
{B\over F}s\cdot\chi_D^i\chi_U^j
-{B\over F}s\cdot\l h_D^iF_{H_U}^j+F_{H_D}^ih_U^j\r
\nonumber
\\&+&
{A^L_{ab}\over F}
\l
l_a^je_b^c\cdot sh_D^i
+l_a^j\chi_D^i\cdot s\tilde{e}_b^c
+s\tilde{l}_a^j\cdot e_b^c\chi_D^i
-sF_{L_a}^j\tilde{e}_b^ch_D^i
-s\tilde{l}_a^jF_{E_b^c}h_D^i
-s\tilde{l}_a^j\tilde{e}_b^cF_{H_D}^i
\r
\nonumber
\\&+&
{A_{ab}^D\over F}
\l
q_a^jd_b^c\cdot sh_D^i
+s\tilde{q}_a^j\cdot d_b^c\chi_D^i
+s\tilde{d}_b^c\cdot q_a^j\chi_D^i
-sF_{Q_a}^j\tilde{d}_b^ch_D^i
-s\tilde{q}_a^jF_{D_b^c}h_D^i
-s\tilde{q}_a^j\tilde{d}_b^cF_{H_D}^i
\r
\nonumber
\\&+&
{A_{ab}^U\over F}
\l
q_a^iu_b^c\cdot sh_U^j
+s\tilde{q}_a^i\cdot u_b^c\chi_U^j
+s\tilde{u}_b^c\cdot q_a^i\chi_U^j
-sF_{Q_a}^i\tilde{u}_b^ch_U^j
-s\tilde{q}_a^iF_{U_b^c}h_U^j
-s\tilde{q}_a^i\tilde{u}_b^cF_{H_U}^j
\r
\nonumber
\\
&+&h.c.\biggr)\;.
\label{S-superpotential}   
\end{eqnarray}
Here we presented only leading order in $1\over F$ terms; the convention for the Levi--Civita tensor is $\epsilon^{0123}=-1$. 

Eliminating auxiliary fields, one gains the low-energy effective Lagrangian 
for the interaction between the components of goldstino supermultiplets and MSSM fields. 
A few remarks are in order. First, being aimed at getting the 
most universal Lagrangian applicable to any supersymmetric extension of the SM, we have to consider only 
interaction terms obtained to the leading order in $1/F$ and to zero order in MSSM coupling constants. Second, 
from the phenomenological point of view the most promising interaction terms are ones which include the minimum {\it new} (additional to SM) fields. In model with R-parity goldstino is R-odd, whereas sgoldstinos are R-even. Consequently, the simplest vertex with goldstino has to contain at least one {\it new} field -- superpartner, while among sgoldstino vertices there is a set containing only SM fields, and only these vertices are of interest. Likewise we reject the interactions with $F$ and $D$ terms, since, being resolved, these auxiliary fields are proportional to SM gauge or Yukawa couplings. Thus, only {\it minimal} {\it sgoldstino-partner-partner} vertices have been included in the presented model:  
\begin{eqnarray} 
{\cal L}_{S}&=&\Biggl(-\sum_{all~gauge\atop fields}{M_\alpha\over4\sqrt{2}F}S\cdot F_{a~\mu\nu}^\alpha F_a^{\alpha~\mu\nu}
\nonumber
\\&-&
\epsilon_{ij}\l
{A^L_{ab}\over \sqrt{2}F}l_a^je_b^c\cdot Sh_D^i
+{A_{ab}^D\over\sqrt{2} F}q_a^jd_b^c\cdot Sh_D^i
+{A_{ab}^U\over\sqrt{2} F}q_a^iu_b^c\cdot Sh_U^j
\r\Biggr)+h.c.\;,
\\
{\cal L}_{P}&=&\Biggl(\sum_{all~gauge\atop fields}{M_\alpha\over 8\sqrt{2}F}P\cdot F_{a~\mu\nu}^\alpha \epsilon^{\mu\nu\lambda\rho}F_{a~\lambda\rho}^\alpha
\nonumber
\\&-&
\epsilon_{ij}\l
i{A^L_{ab}\over\sqrt{2} F}l_a^je_b^c\cdot Ph_D^i
+i{A_{ab}^D\over\sqrt{2} F}q_a^jd_b^c\cdot Ph_D^i
+i{A_{ab}^U\over\sqrt{2} F}q_a^iu_b^c\cdot Ph_U^j
\r\Biggr)+h.c.
\end{eqnarray}
Here we used the replacement~(\ref{definition}). 

Being included into CompHEP, this model should be regarded as an additional option allowed in the framework of CompHEP/SUSY model. Currently accessible versions of this package operate only with real parameters and coupling constants. Likewise the trilinear soft couplings are assumed to be proportional to the corresponding Yukawa couplings, $A_{ab}={\cal A}_{ab}y_{ab}$. The sgoldstino Lagrangian transformed in accordance with these rules reads 
\begin{eqnarray} 
{\cal L}_{S}&=&-\sum_{all~gauge\atop fields}{M_\alpha\over2\sqrt{2}F}S\cdot F_{a~\mu\nu}^\alpha F_a^{\alpha~\mu\nu}
-{{\cal A}^L_{ab}\over\sqrt{2} F}y^L_{ab}\cdot S
\bigl(\epsilon_{ij} l_a^je_b^c h_D^i +h.c.\bigr)
\nonumber
\\&-&
{{\cal A}_{ab}^D\over\sqrt{2} F}y_{ab}^D\cdot S
\bigl( \epsilon_{ij} q_a^jd_b^c h_D^i+h.c.\bigr)
-{{\cal A}_{ab}^U\over\sqrt{2} F}y_{ab}^U\cdot S
\bigl( \epsilon_{ij} q_a^iu_b^ch_U^j+h.c.\bigr)\;,
\label{scalar}
\\
{\cal L}_{P}&=&\sum_{all~gauge\atop fields}{M_\alpha\over 4\sqrt{2}F}P\cdot F_{a~\mu\nu}^\alpha \epsilon^{\mu\nu\lambda\rho}F_{a~\lambda\rho}^\alpha
-i{{\cal A}^L_{ab}\over\sqrt{2} F}y^L_{ab}\cdot P
\bigl(\epsilon_{ij} l_a^je_b^c h_D^i -h.c.\bigr)
\nonumber
\\&-&
i{{\cal A}_{ab}^D\over\sqrt{2} F}y_{ab}^D\cdot P
\bigl( \epsilon_{ij} q_a^jd_b^c h_D^i-h.c.\bigr)
-i{{\cal A}_{ab}^U\over\sqrt{2} F}y_{ab}^U\cdot P
\bigl( \epsilon_{ij} q_a^iu_b^ch_U^j-h.c.\bigr)\;.
\label{pseudoscalar}
\end{eqnarray}

One can check that resulting interaction terms between sgoldstinos and SM gauge bosons coincide with corresponding couplings presented in Ref.~\cite{Perazzi:2000id} except of the terms proportional to 
the parameter $\mu_a$ introduced there. The resulting interaction terms between sgoldstinos and SM fermions coincide with corresponding couplings presented in Ref.\cite{Gorbunov:2000th} up to overall sign.

\subsection{Gravitino--sgoldstino effective Lagrangian}

To obtain the effective Lagrangian for gravitino-sgoldstino sector we again apply the spurion method~\cite{Brignole:1996fn}. It is suitable for our purposes, since the only process we want to describe is the decay of sgoldstinos into gravitinos, i.e., the process with all involved particles being on mass shell. These decays are kinematically allowed in the region of parameter space viable for study at colliders. The relevant terms follow from the effective Lagrangian
\begin{equation}
{\cal L}_\S=\int d\theta^2 d\bar{\theta}^2\l \S^\dag\S-{m_S^2+m_P^2\over 8F^2}\l\S^\dag\S\r^2\r-{m_S^2-m_P^2\over 12F}\l\int d\theta^2 \S^3+h.c.\r\;,
\nonumber
\end{equation}
where we introduced sgoldstino masses. Recalling the definitions~(\ref{spurion}), (\ref{spurion-1}), (\ref{definition}) we obtain the Lagrangian with kinetic and mass terms for goldstino (longitudinal component of gravitino)  and sgoldstinos and the interactions respectable for sgoldstino decays into gravitinos 
\begin{eqnarray}
{\cal L}_{\psi,S,P}&=&i\d_\mu\bar{\psi}\bar{\sigma}^\mu\psi
+\half\d_\mu S\d^\mu S
-\half m_S^2S^2+\half\d_\mu P\d^\mu P-\half m_P^2P^2
\nonumber
\\
&+&{m_S^2\over2\sqrt{2}F}S\l\psi\psi+\bar{\psi}\bar{\psi}\r
-i{m_P^2\over2\sqrt{2}F}P\l\psi\psi-\bar{\psi}\bar{\psi}\r\;.
\label{sgoldstino-gravitino}
\end{eqnarray}
The results for sgoldstino-gravitino coupling constants coincide with corresponding terms presented in Ref.~\cite{Brignole:1996fn} up to overall sign.

\section{Phenomenology of the model}
Till now, there are no experimental evidence for gravitino or sgoldstinos. 
The study of their phenomenology 
results in acquisition of bounds on their coupling constants. 

Let us briefly review the current status of experimental limits on $F$. 

If we ignore sgoldstino (or, equivalently, set $m_{S,P}\to\infty$), then in models with light gravitino 
the current direct bound on the supersymmetry breaking scale is 
obtained from Tevatron, $\sqrt{F}>$217~GeV~\cite{Affolder:2000ef}. 
The upgraded Tevatron may be able
to cover the range of $\sqrt{F}$ up to 
$\sqrt{F}\simeq$290~GeV~\cite{Ambrosanio:1998zf}, while LHC will be capable to
reach $\sqrt{F}\simeq$1.6~TeV~\cite{Brignole:1998me}. 
As we will see, in models with light sgoldstinos, collider 
experiments become more sensitive to the scale of supersymmetry breaking. 

Then, in the model under discussion gravitino is extremely light: $m_{3/2}\lesssim10^{-2}$~eV and may affect 
standard Big Bang Nucleosynthesis as a neutrino-like specie. To get a sufficient dilution the gravitino decoupling temperature must exceed $T\sim 200$~MeV, that yields the limit on the scale of supersymmetry breaking, 
$\sqrt{F}\gtrsim600$~GeV at $m_{\tilde{e}}\simeq100$~GeV~\cite{Moroi:1993mb}. Similar bounds follow from the cooling rate of white dwarfs~\cite{Fukugita:1982eq} and $Z$-boson width~\cite{Dicus:1990vm} in the models where one neutralino is lighter than $Z$-boson. It is worth noting that all these bounds are ensued from the study of the processes with two gravitinos, 
so they are model dependent\footnote{All these bounds were obtained in the framework of linear N=1 supergravity.} and may be invalid in other supersymmetric models.   

Let us turn to the models with light sgoldstinos. 
Note, that all sgoldstino coupling constants introduced in previous sections are
completely determined by MSSM soft terms and the parameter of
supersymmetry breaking $F$, but sgoldstino masses ($m_S,m_P$) remain 
free. Depending on the values of these unknown parameters,
sgoldstinos may show up in different experiments. Phenomenologically
interesting models form four classes: 

1) Sgoldstino masses are of order of electroweak scale, 
$\sqrt{F}\sim1$~TeV --- sgoldstino may be produced 
in collisions of high energy particles at 
colliders~\cite{Perazzi:2000id,Perazzi:2000ty}.

2) Sgoldstino masses $m_S,m_P\sim1$~MeV$\div1$~GeV, 
$\sqrt{F}\sim1$~TeV --- sgoldstinos may emerge in products of rare
decays of mesons~\cite{Dicus:1990su,Gorbunov:2000th}, such as $\Upsilon\to S(P)\gamma$, $J/\psi\to
S(P)\gamma$. 

3) Sgoldstinos in models with flavor violation in the sector of
soft trilinear couplings, $A_{ab}\neq A\delta_{ab}$ --- sgoldstino 
interactions lead to flavor violating processes. Namely, sgoldstinos 
may contribute to 
FCNC (mass difference and/or CP-violation in
the systems of neutral mesons)~\cite{Brignole:2000wd,Gorbunov:2000cz}, 
then, if 
kinematically allowed, they appear
in the product of rare decays, such as 
$t\to cS(P)$~\cite{Gorbunov:2000ht}, $\mu\to eS(P)$,
$K\to\pi S~$\cite{Gorbunov:2000th}, etc.

4) Sgoldstinos are lighter than 1~MeV --- these models may be tested
in low energy experiments~\cite{Gorbunov:2000th}, such as reactor experiments, conversion in magnetic field, etc. Also sgoldstinos may play very important role in astrophysics and 
cosmology~\cite{Nowakowski:1994ag,Grifols:1996hi,Gherghetta:1997zq,Gorbunov:2000th}: they may change the predictions of Big Bang Nucleosynthesis, 
distort CMB spectrum, affect SN explosions and cooling rate of stars, etc.  

There are flavor-conserving and flavor-violating interactions of
sgoldstino fields. 

As concerns flavor-conserving interactions, the strongest bounds arise from astrophysics and
cosmology, that is $\sqrt{F}\gtrsim 10^6$~GeV~\cite{Nowakowski:1994ag,Gorbunov:2000th}, or $m_{3/2}>600$~eV, 
for models with $m_{S(P)}<10$~keV and MSSM soft flavor-conserving 
terms being of the order
of electroweak scale. 
For the intermediate sgoldstino masses (up to a few MeV)
constraints from the study of SN explosions and reactor
experiments lead to $\sqrt{F}\gtrsim300$~TeV~\cite{Gorbunov:2000th}. 
For heavier sgoldstinos, low energy processes 
(such as rare decays of mesons) provide 
limits at the level of $\sqrt{F}\gtrsim500$~GeV~\cite{Gorbunov:2000th}. 

The collider experiments exhibit the same level of sensitivity 
to light sgoldstinos as rare 
meson decays. 
Indeed the 
studies~\cite{Dicus:1989gg,Dicus:1990su,Dicus:1990dy,Dicus:1996ua} 
of the light sgoldstino ($m_{S,P}\lesssim a~few$~MeV) 
phenomenology based on the effective low-energy 
Lagrangian derived from N=1 linear supergravity yield the bounds: 
$\sqrt{F}\gtrsim500$~GeV (combined bound on 
$Z\to S\bar{f}f,P\bar{f}f$~\cite{Dicus:1990dy}; combined bound on 
$e^+e^-\to\gamma S,\gamma P$~\cite{Dicus:1990su}) at $M_{soft}\sim100$~GeV, 
$\sqrt{F}\gtrsim1$~TeV~\cite{Dicus:1996ua} 
(combined bound on $p\bar{p}\to gS,gP$) at $M_3\simeq500$~GeV. 
Searches for heavier 
sgoldstino at colliders, though exploiting another technique, results in 
similar bounds on the scale of supersymmetry breaking.  
Most powerful among the operating machines, LEP and
Tevatron, give a constraint of the order of 1~TeV on supersymmetry breaking
scale in models with light sgoldstinos. Indeed, the analysis carried out by 
DELPHI Collaboration~\cite{Abreu:2000ij} yields the limit 
$\sqrt{F}>500\div200$~GeV at sgoldstino masses $m_{S,P}=10\div150$~GeV and 
$M_{soft}\sim100$~GeV. 
The constraint depends on the MSSM soft breaking parameters. In particular, 
it is stronger by about a hundred GeV in the model with degenerate gauginos. 
At Tevatron, a few events in
$p\bar{p}\to S\gamma(Z)$ channel, and about $10^4$ events in $p\bar{p}\to S$
channel would be produced at $\sqrt{F}=1$~TeV and $M_{soft}\sim100$~GeV 
for integrated luminosity 
${\cal L}=100$~pb$^{-1}$ and sgoldstino mass of order 
of 100~GeV~\cite{Perazzi:2000ty}. 
This gives rise to a possibility to detect
sgoldstino, if it decays inside the detector into photons and $\sqrt{F}$
is not larger than $1.5\div2$~TeV. 

The flavor-asymmetric processes are generally very sensitive to
light sgoldstino. Namely, with flavor-changing
off-diagonal entries in squark (slepton) squared mass matrix close to
the current bounds, direct
measurements of decays of mesons (leptons) provide very strong bounds,
up to $\sqrt{F}\gtrsim900(15000)$~TeV~\cite{Gorbunov:2000th} (valid at $m_{S}\lesssim5(0.34)$~GeV), 
which is much higher than bounds expected from future
colliders. Sgoldstinos may appear in top-quark decay, that could 
help to probe the scale of supersymmetry breaking at LHC 
up to about 10~TeV~\cite{Gorbunov:2000ht}. 
The possible contributions of sgoldstinos with $m_{S,P}=100$~GeV to mass differences in the systems of neutral mesons yield the bounds 
$\sqrt{F}\gtrsim10$~TeV~\cite{Gorbunov:2000cz}. If off-diagonal entries are small, the limits on $\sqrt{F}$ become 
weaker: they scale as square root of the corresponding 
off-diagonal elements. 

Among the laboratory processes, the most sensitive to very light sgoldstinos
are propagation of laser beam in magnetic field and reactor
experiments. For heavier sgoldstinos 
the most sensitive processes to flavor-conserving sgoldstino couplings 
are $\Upsilon$ decays and associated sgoldstino production in collisions of 
high energy particles; charged kaon decays are most sensitive to 
flavor-violating sgoldstino couplings. 
For sgoldstino masses of the order of electroweak scale the most sensitive process to flavor-conserving sgoldstino couplings is sgoldstino resonant production at hadron colliders, while the most sensitive to flavor-violating couplings is measurement of mass differences in the systems of neutral mesons.

\section{CompHEP notations}
In the CompHEP package gravitino and sgoldstinos ($\psi$, $S$, $P$) are denoted as~ ${\tt \tilde{}\!~g}$, ${\tt sS}$ and ${\tt sP}$, respectively. Sgoldstino masses $m_S$ and $m_P$ are referred as ${\tt MsS}$ and ${\tt MsP}$; gravitino is massless. The scale of supersymmetry breaking in the entire theory, $\sqrt{F}$, corresponds to ${\tt sqF}$.  
The Feynman rules for the Lagrangians~(\ref{goldstino}), 
(\ref{scalar}), (\ref{pseudoscalar}), (\ref{sgoldstino-gravitino}) have been obtained by making use of LanHEP package~\cite{Semenov:eb}. They have been incorporated into CompHEP, as well as aforementioned new parameters with default values ${\tt MsS}={\tt MsP}=200$~GeV, ${\tt sqF}=3$~TeV.\footnote{These values may be changed by user that is a standard CompHEP menu option.} Subsequent checks gave results consistent with ones known in literature.   

\section{Concluding remarks}

We described the model with very light gravitino and light sgoldstinos, 
which has been included into CompHEP/SUSY package. 
This package may be used in order to calculate a rate of 
any tree-level process with one gravitino or sgoldstino being on mass shell. This includes: 
\begin{itemize}
\item decay of a superpartner into partner and gravitino 
\item sgoldstino decays
\item gravitino inelastic scattering off MSSM particles into MSSM particles 
\item sgoldstino inelastic scattering off SM particles into MSSM particles 
\item gravitino (sgoldstino) production in collisions of MSSM particles.
\end{itemize}
The effective Lagrangian incorporated into CompHEP/SUSY is universal, so the package may be used to study the phenomenology in any supersymmetric extension of the Standard Model. 
The only interaction terms to the leading order 
in $1/F$ and to zero order in MSSM gauge and Yukawa coupling constants, 
relevant for gravitino and sgoldstino phenomenology,  
are very terms included into this package. 

Note, that direct independent measurement of MSSM soft supersymmetry breaking
terms and gravitino or/and sgoldstino couplings gives the unique possibility to estimate the scale of supersymmetry breaking $\sqrt{F}$.  

As a final check for gravitino couplings we compared the analytical results (from CompHEP output) for all decay rates of the kind {\it superpartner}$\to${\it partner+gravitino} with corresponding formulae presented in Ref.~\cite{Ambrosanio:1996jn} and found no difference. To make sure that sgoldstino vertices are correct we compared the analytical results (from CompHEP output) for all sgoldstino decay rates with corresponding formulae presented in 
Refs.~\cite{Perazzi:2000id,Gorbunov:2000th} The results coincide up to terms proportional to additional parameter $\mu_a$ introduced in Ref.~\cite{Perazzi:2000id}. To check the numerical calculations carrying out by the package, we have reproduced with good accuracy all results for sgoldstino production in various channels presented in Refs.~\cite{Perazzi:2000id,Perazzi:2000ty}. 

There are several remarks which may be helpful for the user who are going to work with this package.  

It is worth recalling that we discussed the model with R-parity being conserved. Gravitino is R-odd, consequently, it might appear in collision of SM particles only accompanied by another R-odd particle. Moreover, gravitino is stable the lightest supersymmetric particle --- LSP. Sgoldstinos are R-even, therefore, their production may be not accompanied by production of 
superpartners. Sgoldstinos may decay into SM particles, and since two-photon channel is always open, they are unstable. 

The second remark concerns the treatment of sgoldstino coupling constants to be proportional to the MSSM soft breaking terms. In fact, the Lagrangians~(\ref{scalar}), (\ref{pseudoscalar}) describe sgoldstino interactions at
the superpartner scale. Sgoldstino coupling constants at other energies may
be obtained by making use of renormalization group evolution. As an example, 
for the gluonic operator one has
\begin{equation}
G_{\mu\nu}^2(M_3)=G_{\mu\nu}^2(\mu){\beta(\alpha_s(\mu))\over\alpha_s(\mu)}
{\alpha_s(M_3)\over\beta(\alpha_s(M_3))}\;,
\nonumber
\end{equation} 
where $\beta(\alpha_s)$ is QCD $\beta$-function. These corrections (renormalization group enhancement) are neglected in the package, but may be taken into account plainly by making use of appropriate rescaling. 

Then, in CompHEP all Yukawa coupling constants are determined as the ratios of fermion pole masses to the corresponding Higgs vacuum expectation value. In order to estimate sgoldstino branching ratios into fermions one should take into account the difference between running Yukawa constants and Yukawas used in CompHEP. 
 
The forth remark is that sgoldstino couplings to superpartners are not included into this package. If open (allowed kinematically), new sgoldstino 
decay channels distort the pattern of sgoldstino branching ratios calculated in CompHEP. In a given model the level of distortion depends on the set of soft supersymmetry breaking masses and sgoldstino masses and effective couplings. 

Finally, we discuss the possible directions to improve this package. The first interesting possibility is to take into account CP- and flavor-violating sgoldstino couplings. This option will be allowed, if complex coupling constants and flavor violating couplings become available in CompHEP. Then, sgoldstino couplings to superpartners may be also included. In the spurion approach corresponding coupling constants are not completely determined by MSSM soft breaking terms, but may depend on new parameters originated from the hidden sector. Similar situation takes place with sgoldstino interaction terms proportional to MSSM gauge or Yukawa couplings. For instance, the account of $F$ and $D$ auxiliary fields (see Eqs.~(\ref{S-kahler}), (\ref{S-gauge}), (\ref{S-superpotential})) results, in particular, in sgoldstino-higgs mixing with coefficients depending also on new parameters from hidden sector. The model independence of these coefficients will be clarified elsewhere.  

\section{Acknowledgments}

The authors are indebted to V.~Ilyin 
for helpful discussions on different subjects related to
this work. D.G. thanks the LAPTH high energy 
physics group for their warm hospitality while this work has been done. 
This work is supported in part by RFBR grant 01-02-16710, and is
performed in the framework of CPP Collaboration. 
The work of D.G.\ is supported in part by RFBR grants 99-02-18410a, 
01-02-06035, by CPG and SSLSS grant 00-15-96626, 
by CRDF award RP1-2103, 
and by Swiss National Science Foundation, project No.~7SUPJ062239. 
The work of A.S.\ is supported in part by CERN-INTAS grant 99-0377. 


\end{document}